\documentclass[aps,pre,10pt,twocolumn]{revtex4-1}
\usepackage{graphicx}
\usepackage{epstopdf}
\usepackage{amsmath}
\usepackage{amssymb}
\usepackage[utf8]{inputenc}
\usepackage{graphicx,float,hyperref}
\usepackage{natbib}
\usepackage{graphicx,latexsym} 
\DeclareMathOperator{\sech}{sech}

\begin{document}


\title{Performance bounds of non-adiabatic quantum harmonic Otto engine and refrigerator under a squeezed thermal reservoir} 

\author{Varinder Singh}
\email{vsingh@ku.edu.tr}
\affiliation{Department of Physics, Ko\c{c} University, 34450 Sar\i{}yer, Istanbul Turkey}
\author{\"{O}zg\"{u}r E. M\"{u}stecapl{\i}o\u{g}lu}
\email{omustecap@ku.edu.tr}
\affiliation{Department of Physics, Ko\c{c} University, 34450 Sar\i{}yer, Istanbul Turkey}
%


\begin{abstract}
We analyze the performance of a quantum Otto cycle, employing time-dependent harmonic oscillator 
as the working fluid undergoing sudden expansion and compression strokes during the adiabatic stages, 
coupled to a squeezed reservoir. First, we show that the maximum efficiency that our engine can 
achieve is 1/2 only, which is in contrast with the earlier studies claiming unit efficiency under the 
effect of squeezed reservoir. Then, we obtain analytic expressions for the upper bound on the 
efficiency as well as on the coefficient of performance of the Otto cycle. The obtained bounds are 
independent of the parameters of the system and  depends on the reservoir parameters only. 
Additionally, with hot squeezed thermal bath, we obtain analytic expression for the efficiency at 
maximum work which satisfies the derived upper bound. Further, in the presence of squeezing in the 
cold reservoir, we specify an operational regime for the Otto refrigerator otherwise forbidden in 
the standard case. 
\end{abstract}

\pacs{03.67.Lx, 03.67.Bg}

\maketitle 

%
\section{Introduction}
The concept of Carnot efficiency ($\eta_C$) is one of the most important results in physics, which led to the formulation of the second
law of thermodynamics \cite{Kondepudi}. It puts a theoretical
upper bound on the efficiency of all macroscopic heat engines working between two thermal reservoirs at different temperatures. However, with the rise of quantum thermodynamics \cite{SV2016,Mahler,DeffnerBook,AlickiKosloff}, many studies have showed that this sacred bound may be surpassed by quantum heat machines  exploiting exotic quantum resources such as quantum coherence 
\cite{Scully2001,ScullyAgarwal2003,OzgurQuantumFuel,Skrzypczyk2014}, quantum correlations \cite{MNBera,Park2013,Brunner2014,Marti2015,OzgurCorrelation}, squeezed reservoirs \cite{Lutz2014,Huang2012,Rezek2017,Bijay2017,Manzano2016,Xiao2018,Long2015,Klaers2017,Alonso2014,Assis,Wang2019}, among others. In such cases, the  second law of thermodynamics has to be modified to account for the quantum effects, and the notion of generalized Carnot bound is introduced which is always satisfied \cite{MNBera,Niedenzu2018,Obinna2014,Lutz2014}. In this context, different theoretical studies have been
carried out to study the implications of work extraction when quantum heat machines are coupled to nonequilibrium stationary reservoirs \cite{Niedenzu2016,Alicki2015,Obinna2014,Alicki2014,Ghosh2018}. In particular, it is instructive to look into the working of heat machines coupled to squeezed thermal reservoirs. The use of squeezed thermal reservoir allows us to extract work from a single reservoir \cite{Manzano2016}, operate thermal devices beyond Carnot bound \cite{Klaers2017,Lutz2014,Manzano2016,Long2015}, define multiple operational  regimes \cite{Manzano2016,Niedenzu2016}  otherwise impossible for the standard case with two thermal reservoirs. Moreover, in Ref. \cite{Manzano2018}, the idea of treating squeezed thermal reservoir as a generalized equilibrium reservoir is explored.  Recently, a nanomechanical engine consisting of a vibrating nanobeam coupled to squeezed thermal noise, operating beyond the standard Carnot efficiency, is  realized experimentally \cite{Klaers2017}. 

Over the past few years, there have been increasing interest in investigating the performance of a quantum Otto cycle
\cite{Quan2007,Kieu2004,Rezek2006,Lutz2012,GJ2011,Suman2,Peterson2019}, based 
on a time-dependent harmonic oscillator as the working fluid, coupled to squeezed thermal baths \cite{Lutz2014,Long2015,Manzano2016,Xiao2018,Klaers2017}. Due to its simplicity, 
harmonic quantum Otto cycle (HQOC) serves as a paradigm model for quantum thermal devices. It consists of two adiabatic  branches during which the frequency of the oscillator is varied, and two isochoric branches during which the system exchanges heat with the thermal baths at constant frequency. 
Ro\ss{}nagel and coauthors 
optimized the work output of a HQOC in the presence of hot squeezed thermal bath and obtained generalized version of Curzon-Ahlborn efficiency \cite{Lutz2014}. Manzano et. al studied a modified version of HQOC  and discussed the effect of squeezed hot bath in different operational regimes \cite{Manzano2016}. Extending the analysis to the quantum refrigerators, Long and Liu optimized the performance 
of a HQOC in contact with low temperature squeezed thermal bath and concluded that the coefficient of performance (COP) can be enhanced by squeezing \cite{Long2015}.

With the exception of Refs. \cite{Xiao2018,Assis}, all the above-mentioned studies involving squeezed reservoirs 
are confined to the study of quasi-static Otto cycle in which adiabatic steps are performed quasi-statically, thus 
producing vanishing power output.  
In this work, we fill this gap by confining our focus to the highly non-adiabatic (dissipative) regime corresponding to the sudden switch of frequencies (sudden compression/expansion strokes) during the adiabatic stages of the Otto cycle. 
We obtain analytic expressions for the upper bounds on the efficiency and COP of
the HQOC coupled to  a squeezed thermal reservoir.

\begin{figure}[ht]
 \begin{center}
\includegraphics[width=8.6cm]{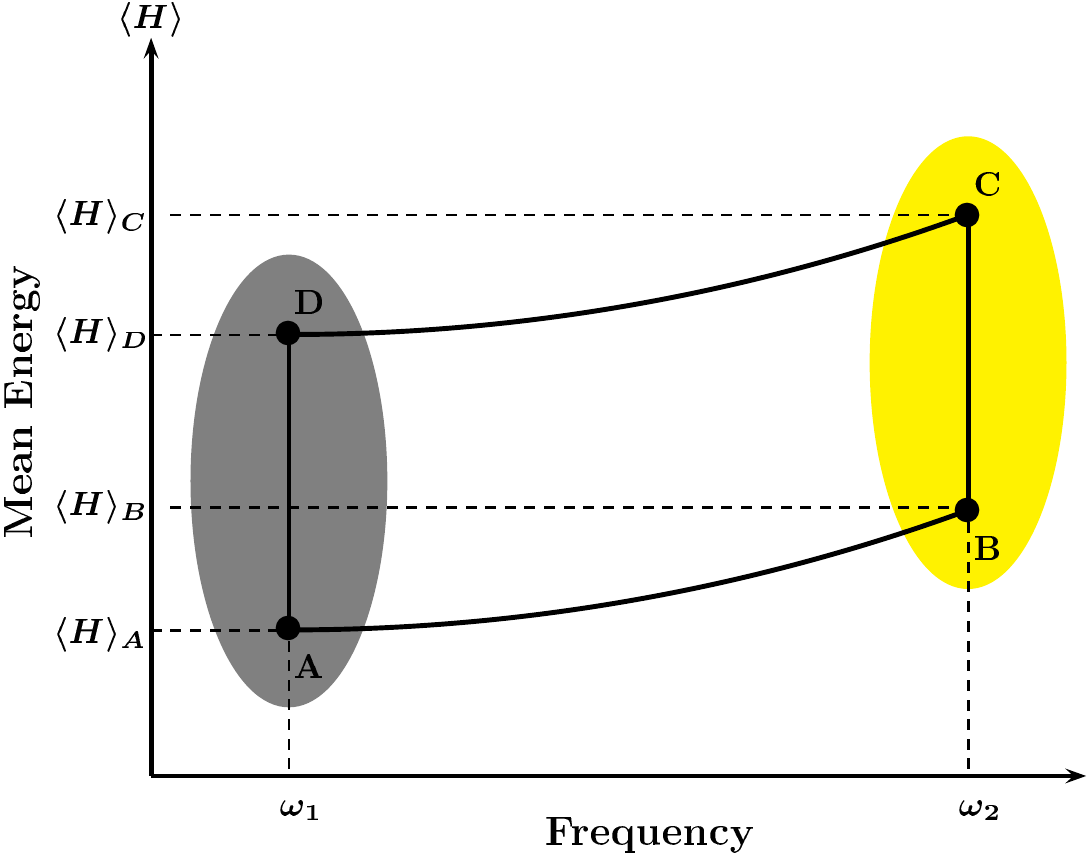}
 \end{center}
\caption{Model of quantum Otto Cycle employing time-dependent harmonic oscillator as the working fluid.}
\end{figure}
The paper is organized as follows. In Sec. II, we discuss the model of HQOC coupled to a hot squeezed thermal reservoir. In Sec. III, we obtain analytic expression for the upper bound on the efficiency of the engine operating in the sudden switch limit. We also obtain analytic expression for the efficiency at maximum work and compare it with the derived upper bound. In Sec. IV, we repeat our analysis for the Otto refrigerator coupled to a cold squeezed reservoir and obtain upper bound on the COP of the refrigerator. We conclude in Sec. V.

%

\section{Quantum Otto cycle with squeezed reservoir}
We consider quantum Otto cycle of a time-dependent harmonic oscillator coupled to a hot squeezed thermal bath 
while cold bath is still purely thermal in nature. It consists of four stages: two adiabatic and two isochoric. 
These processes occur in the following order \cite{Lutz2012,LutzEPL}:
(1) Adiabatic  compression $A \longrightarrow B$:  To begin with, the system is at 
inverse temperature $\beta_1$. The system is isolated and frequency of the oscillator is increased from 
$\omega_1$ to $\omega_2$. Work is done on the system in this stage. The evolution is unitary and von Neumann 
entropy of the system remains constant.
(2) Hot isochore $B\longrightarrow C$: During this stage, the oscillator is coupled to the squeezed thermal heat reservoir at 
inverse temperature $\beta_2$ at fixed frequency ($\omega_2$) and allowed to thermalize. No work is done in this stage, only heat exchange between the system and reservoir takes place. 
After the completion of the hot 
isochoric stage, the system relaxes to  a nondisplaced squeezed thermal state \cite{Kim1989,Marian1993} with mean photon number
$\langle n(\beta_2,r)\rangle = \langle n\rangle+(2\langle n\rangle+1)\sinh^2r$, where $r$ is the squeezing 
parameter and $\langle n\rangle=1/(e^{\beta_2\omega_2}-1)$ is the thermal occupation number (we  have set $\hbar=k_B=1$ for simplicity). 
%
(3) Adiabatic expansion $C \longrightarrow D$: The system is isolated and the frequency of the oscillator is unitarily 
decreased back to its initial value $\omega_1$. Work is done by the system in this stage.   
(4) Cold isochore $D\longrightarrow A$: To bring back the working fluid to its initial state, the system is coupled to the cold reservoir at inverse temperature $\beta_1$ ($\beta_1>\beta_2$), and allowed to relax back to the initial thermal state $A$.

The average  energies $\langle H \rangle$ of the oscillator at the four stages of the cycle
read as follows \cite{Lutz2014}:

\begin{equation}
\langle H\rangle_A =\frac{ \omega_1}{2}\text{coth}\Big(\frac{\beta_1  \omega_1}{2}\Big) ,
\end{equation}
\vspace{1mm}
\begin{equation}
\langle H\rangle_B =\frac{ \omega_2}{2} \lambda \text{coth}\Big(\frac{\beta_1  \omega_1}{2}\Big) ,
\end{equation}
\begin{equation}
\langle H_C\rangle = \frac{\omega_2}{2}\coth\left(\frac{\beta_2\omega_2}{2}\right)\Delta H(r),
\end{equation}
\begin{equation}
\langle H_D\rangle = \frac{\omega_1}{2}\lambda \coth\left(\frac{\beta_2\omega_2}{2}\right)\Delta H(r),
\end{equation}
where $\Delta H(r)=\langle n(\beta_2,r)\rangle/\langle n\rangle=1+(2+1/\langle n\rangle)\sinh^2r$ 
reflects the effect of the squeezed hot thermal bath on the mean energy of the oscillator, $\lambda$  
is the dimensionless adiabaticity parameter \cite{Husimi}. For the  adiabatic process, $\lambda=1$; 
for non-adiabatic expansion and compression strokes, $\lambda>1$. 
The expression for  mean heat exchanged during the hot and cold isochores can be evaluated, respectively, as follows:
 \begin{eqnarray}
  \langle Q_2 \rangle &=& \langle H \rangle_C-\langle H \rangle_B \nonumber
  \\
  &=& \frac{ \omega_2}{2}\Big[\Delta H(r)\text{coth}\Big(\frac{\beta_2 \omega_2}{2}\Big)-\lambda\text{coth}\Big(\frac{\beta_1  \omega_1}{2}\Big) \Big] ,\nonumber
  \\ 
  \label{heat2}
\\
  \langle Q_4 \rangle &=& \langle H \rangle_A-\langle H \rangle_D \nonumber
  \\
  &=& \frac{  \omega_1}{2}\Big[\text{coth}\Big(\frac{\beta_1 \omega_1}{2}\Big)-\lambda\Delta H(r)\text{coth}\Big(\frac{\beta_2\omega_2}{2}\Big) \Big].\nonumber
  \\   
  \label{heat4}   
 \end{eqnarray}
Here, we are employing a sign convention in which heat absorbed (rejected) from (to) the
reservoir is positive (negative) and work done on (by) the system is positive (negative).

Since after one complete cycle, the working fluid comes back to its initial state, the extracted work in one 
complete cycle is given by, $\langle W_{\rm ext}\rangle=\langle Q_2\rangle+\langle Q_4\rangle>0$. In this work, we are 
interested in the sudden switch case for which $\lambda=(\omega_1^2+\omega_2^2)/2 \omega_1 \omega_2$ \cite{Husimi,Deffner2008,Deffner2010}. Substituting above expression for $\lambda$  in Eqs. (\ref{heat2})
and (\ref{heat4}), we obtain the following  expressions for the extracted work, $\langle W_{\rm ext}\rangle$, and efficiency, 
$\eta=\langle W_{\rm ext}\rangle/\langle Q_2\rangle$, of the engine, respectively:
%
%
\begin{widetext}
\begin{equation}
 \langle W_{\rm ext}\rangle = \langle Q_2\rangle+\langle Q_4\rangle = \frac{\omega_2^2-\omega_1^2}{4\omega_1\omega_2} \left[ \omega_1\Delta H(r)\coth \left(\frac{\beta_2\omega_2}{2}\right)
-\omega_2 \coth\left(\frac{\beta_1\omega_1}{2}\right) \right], \label{workgen}
\end{equation}
\begin{equation}
\eta  = \frac{\langle W_{\rm ext}\rangle}{\langle Q_2 \rangle} = \left[\frac{2}{1-\frac{\omega_1^2}{\omega_2^2}} +\frac{1}
{\frac{\omega_1}{\omega_2}\Delta H(r)\coth\big(\frac{\beta_2\omega_2}{2}\big)\tanh\big(\frac{\beta_1\omega_1}{2}\big)-1}\right]^{-1}  . \label{effgen}
\end{equation}
\end{widetext}
Now, efficiency, $\eta$, can attain its maximum when the expression inside the square bracket attains its  minimum value. The minimum value of the first term can be inferred as follows: $Min[A_1]\equiv Min[2/(1-\omega_1^2/\omega_2^2)]=2$, as $\omega_1<\omega_2$ for the engine operation. Similarly,  $Min[A_2]\equiv Min[1/\{\frac{\omega_1}{\omega_2}\Delta H(r)\coth(\beta_2\omega_2/2)\tanh(\beta_1\omega_1/2)-1\}]=0$,   which can be inferred from the positive work condition, $\langle W_{\rm ext}\rangle>0$
[see Eq. (\ref{workgen})], which implies that 
$\frac{\omega_1}{\omega_2}\Delta H(r)\coth(\beta_2\omega_2/2)\tanh(\beta_1\omega_1/2)>1$. Thus, we can conclude that 
the efficiency of a harmonic quantum Otto engine, operating in the sudden switch limit, is bounded from above by one-half the unit value, i.e., 
\begin{equation}
\eta \leq \frac{1}{2}\equiv \eta_{\rm max}.
\end{equation}
This is our first main result. The result is very interesting as it implies that even in the presence of very
very large squeezing ($r\rightarrow\infty$), the efficiency of the engine can never surpass 1/2. This is in contrast 
with the previous studies, valid for the quasi-static regime, implying that the thermal engine fueled by a hot squeezed 
thermal reservoir asymptotically attains unit efficiency for large squeezing parameter ($r>>1$) \cite{Lutz2014,Manzano2016,Niedenzu2018}. 
We attribute this to the highly frictional nature of the sudden switch regime as explained below. In the 
sudden switch regime, the sudden quench of the frequency of the harmonic oscillator induces non-adiabatic 
transitions between its energy levels, thereby causing the system to develop coherence in the
energy frame. In such a case, the energy entropy increases and an additional
parasitic internal energy is stored in the working medium. The additional energy
corresponds to the waste (or excess) heat which is dissipated to the heat reservoirs
during the proceeding isochoric stages  of the cycle \cite{Rezek2010,Plastina2014}. This limits the performance of the
device under consideration. 
\section{Upper bound on the efficiency}
In order to obtain analytic expression in closed form for the efficiency, we will work in the 
high-temperature regime \cite{Kosloff1984,UzdinEPL,VJ2019}. In this regime, we set
$\coth(\beta_i\omega_i/2)\approx 2/(\beta_i\omega_i)$ ($i=1, 2$) and $\Delta H(r)=\cosh(2r)$. Then, the expressions for the 
extracted work $\langle W_{\rm ext}\rangle$ [Eq. (\ref{workgen})] and the efficiency [Eq. (\ref{effgen})]
take the following forms:
\begin{eqnarray}
\langle W_{\rm ext}\rangle &=& \frac{(1-z^2) \left[z^2 \cosh (2 r)-\tau \right]}{2z^2\beta_2}, \label{worksq}
\\
\eta &=& \frac{(z^2-1)[z^2\cosh (2r) - \tau]}{\tau - z^2[2\cosh(2r)-\tau]}, \label{effsss}
\end{eqnarray}
where we have defined $z=\omega_1/\omega_2$ and $\tau=\beta_2/\beta_1=1-\eta_C$.
From Eq. (\ref{worksq}), the positive work condition, $\langle W_{\rm ext}\rangle>0$, implies that
\begin{equation}
z^2\cosh(2r) > 1-\eta_C.  \label{PWC}
\end{equation}
Using the expression for efficiency in Eq. (\ref{effsss}), $z^2$ can be written in terms of $\eta$ and $\eta_C$, and is given by
\begin{widetext}
\begin{equation}
z ^2 = \frac{1}{2} 
\left\{ 
(1-\eta_c)(1+\eta) + (1-2\eta)\cosh(2r)
-
\sqrt{\big[(1-\eta_c)(1+\eta) + (1-2\eta)\cosh(2r)\big]^2 - 4(1-\eta_c)(1+\eta)\cosh(2r) }
\right\} .
\end{equation}
Using the above expression for $z$ in Eq. (\ref{PWC}), we obtain following upper bound on the efficiency of the engine:
\begin{equation}
\eta < \frac{\left[1-\eta _C-\cosh (2 r) \right] \left[-1+\eta _C-2 \cosh (2 r)+2   \sqrt{2\left[1-\eta _C\right] 
\cosh (2 r)}\right]}{\left[1-\eta _C-2 \cosh (2 r)\right]{}^2} \equiv \eta_{\rm up}. \label{etasq}
\end{equation}
\end{widetext}
This is our second main result. Notice that the above derived bound is independent of the  parameters of the 
model under consideration and depends on the reservoir parameters $r$ and $\eta_C$ (or $\tau$) only. 
For $r\rightarrow \infty$, $\eta_{\rm up}\rightarrow 1/2$, which reconfirms our earlier result [Eq. (9)] that 
the maximum efficiency that our  engine can attain is one-half the unit efficiency; it never reaches unit efficiency 
unlike the engines operating in the quasi-static regime \cite{Lutz2014,Manzano2016,Niedenzu2018}.

Further, we derive analytic expression for the efficiency at maximum work  by optimizing Eq. (\ref{worksq})
with respect to $z$, and it is given by:
\begin{equation}
\eta_{\rm MW} = \frac{1-\sqrt{(1-\eta_C)\sech(2r)}}{2+\sqrt{(1-\eta_C)\sech(2r)}}. \label{etaRKsq}
\end{equation}
We have plotted Eqs. (\ref{etasq}) 
and (\ref{etaRKsq})  in Fig. 2 as a function of $r$ for different fixed values of Carnot efficiency $\eta_C$. 
For the given values of $\eta_C$ smaller than 1/2, both $\eta_{\rm up}$ (solid red and blue curves)
and $\eta_{\rm MW}$ (dashed red and blue curves) can surpass corresponding Carnot efficiency 
(dotted curves with same color) for some value of squeezing parameter $r$ and approach 1/2 for relatively larger 
values of $r\,(r>5)$. From the inset of Fig. 2, it is clear that $\eta_{\rm MW}$ always lies below 
$\eta_{\rm up}$, which should be the case as for the given temperature ratio ($\eta_C$), $\eta_{\rm up}$ is the upper
bound on the efficiency.    

One more comment is in order here. Although, for given values of $\eta_C$ ($\eta_C<1/2$),  $\eta_{\rm up}$ and 
$\eta_{\rm MW}$ may surpass standard Carnot efficiency, 
they can never surpass generalized Carnot efficiency (not shown in Fig. 2) \cite{Alicki2014,Lutz2014}, 
\begin{equation}
\eta^{\rm gen}_C = 1 - \frac{\beta_2}{\beta_1\cosh (2r)} \equiv 1 - \frac{T_1}{T_2 \cosh (2r)},\label{Carnotgen}
\end{equation}
which follows from the second law of thermodynamics applied to the nonequilibrium situations \cite{Obinna2014}.
The concept of generalized Carnot efficiency can be understood as follows. We can always assign a frequency 
dependent local temperature to a squeezed thermal reservoir  characterized by its genuine temperature $T$ 
and squeezing parameter $r$ \cite{Alicki2014,Alicki2015}. The expression for this frequency dependent 
local temperature can be obtained from the following relation \cite{Alicki2014,Alicki2015}:
\begin{equation}
\exp\left(-\frac{\omega}{T(\omega,r)}\right) = \frac{\langle n\rangle + (2\langle n\rangle + 1)\sinh^2r}
{1+\langle n\rangle + (2\langle n\rangle + 1)\sinh^2r} \label{effectiveTemp}.
\end{equation}
\begin{figure}[ht]
 \begin{center}
\includegraphics[width=8.6cm]{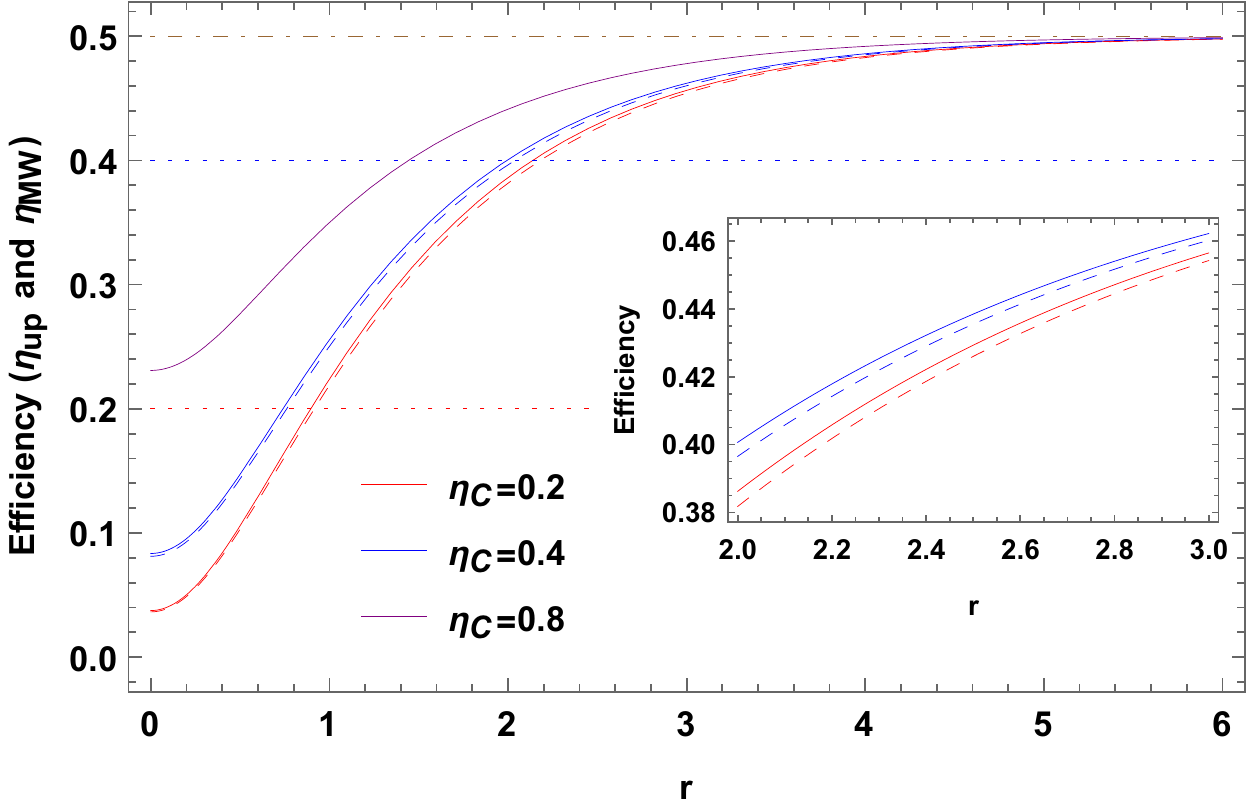}
 \end{center}
\caption{Plots of  $\eta_{\rm up}$ [Eq. (\ref{etasq})] and $\eta_{\rm MW}$ [Eq. (\ref{etaRKsq})] as a function of squeezing parameter $r$. Solid red and blue curves represent $\eta_{\rm up}$ for $\eta_C=0.2$  and $\eta_C=0.4$, respectively. Dashed  curves in the corresponding color represent $\eta_{\rm MW}$.
Dotted red and blue curves denote the standard Carnot efficiency at values $\eta_C=0.2$ and $\eta_C=0.4$, respectively. 
Solid purple curve represent  $\eta_{\rm up}$ for $\eta_C=0.8$, and shows that for the given value of 
$\eta_C>1/2$,  $\eta_{\rm up}$ can never surpass Carnot efficiency $\eta_C$ even in the presence of  
very large squeezing. For the better resolution, in the inset, we have plotted  $\eta_{\rm up}$ and  $\eta_{\rm MW}$ 
for smaller range of $r$. It shows that for the same value of squeezing parameter $r$,  $\eta_{\rm MW}$ always 
lies below $\eta_{\rm up}$.}
\end{figure}
In the high-temperature limit, the effective temperature of the squeezed hot bath reads as,
\begin{equation}
T^{\rm eff}_2(r) = T_2 (1+2\sinh^2r) =T_2 \cosh(2r) \label{effectiveTemp2}.
\end{equation}
Hence, for positive values of $r$, engine may be assumed to be operating between temperatures $T_1$
and $T^{\rm eff}_2(r)$. The  actual (generalized) Carnot efficiency should then be given by Eq. (\ref{Carnotgen}).

Finally, we discuss the special case when $r\rightarrow 0$. This corresponds to the case in which our 
harmonic quantum  engine is working between two purely thermal reservoirs. Thus, for $r\rightarrow 0$, 
Eqs. (\ref{etasq}) and (\ref{etaRKsq}) reduce to the following forms, respectively:
\begin{equation}
\eta <\frac{[3-2\sqrt{2(1-\eta_C)}-\eta_C]\eta_C}{(1+\eta_C)^2} \equiv \eta^{\rm th}_{\rm up}, \label{effmax}
\end{equation}
\vspace{-5mm}
\begin{equation}
\eta_{\rm RK} = \frac{1-\sqrt{1-\eta_C}}{2+\sqrt{1-\eta_C}}. \label{etaRK}
\end{equation}
The above bound, $\eta^{\rm th}_{\rm up}$, is much tighter than the classical Carnot bound, even tighter 
than $\eta_C/2$ (see Fig. 3). Eq. (\ref{etaRK}), which we derived as a special case of our more general result
Eq. (\ref{etasq}), was first derived by Rezek and Kosloff (RK) for the optimization of a 
harmonic quantum Otto engine undergoing sudden switch of frequencies in the adiabatic stages
 \cite{Rezek2006}. Again, it is clear from Fig. 3 that 
$\eta_{\rm RK}$ (dashed red curve) always lies below $\eta^{\rm th}_{\rm up}$ (solid blue curve), which should 
be the case. 
\begin{figure}[ht]
 \begin{center}
\includegraphics[width=8.6cm]{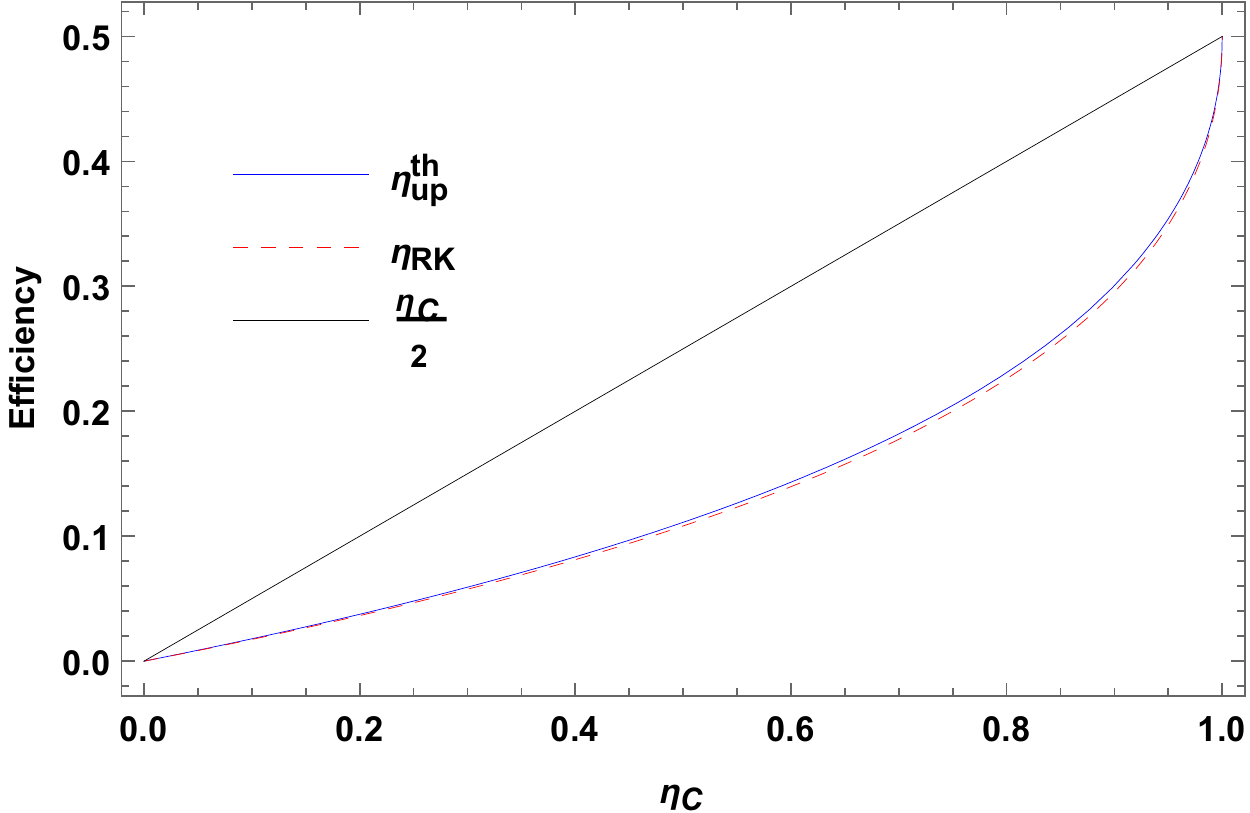}
 \end{center}
\caption{Plots of $\eta^{\rm th}_{\rm up}$ [Eq. (\ref{effmax})], $\eta_{\rm RK}$ [Eq. (\ref{etaRK})]  versus Carnot efficiency. We can see that $\eta_{\rm RK}$ (dashed red curve) lies below $\eta^{\rm th}_{\rm up}$ (solid blue curve). Both are bounded above by half the Carnot efficiency, $\eta_C/2$. }
\end{figure}

\section{Upper bound on the coefficient of performance}
Here, we discuss the operation of QHOC  as a refrigerator. In the refrigeration process, heat is absorbed from the cold bath,
$\langle Q_4\rangle>0$, and dumped into the hot bath, $\langle Q_2\rangle<0$. The net work investigated in the system 
is positive, $\langle W_{\rm in}\rangle=-(\langle Q_2\rangle + \langle Q_4\rangle)>0$. Here, we will first discuss the 
case when refrigerator is coupled to two purely thermal reservoirs. We follow the same procedure as done for the heat engine in Sec. III. Since the calculations are straight forward, we merely present our results here. For the refrigerator running between two purely thermal reservoirs, positive cooling condition, $\langle Q_4\rangle>0$, implies that
\begin{equation}
\zeta_C > 1 \quad \text{\rm and} \quad  \zeta \leq 1+3\zeta_C - 2\sqrt{2\zeta(1+\zeta_C)} \equiv \zeta^{\rm th}_{\rm up}, \label{maxcop}
\end{equation}
where $\zeta=\omega_1/(\omega_2-\omega_1)$ and $\zeta_C=\beta_2/(\beta_1-\beta_2)$ are the COP and Carnot COP, 
respectively. The condition $\zeta_C > 1$ implies that $\tau>1/2$, which in turns implies that cold reservoir
cannot be cooled below the temperature $T_2/2$, thus putting a restriction to the operation of the refrigerator 
operating in sudden-switch regime. The upper bound $\zeta_{\rm up}$ derived here is independent of the 
parameters of the system and depends on ratio of the reservoir temperatures only, which makes it quite general 
in nature. Similar to the heat engine case, the obtained upper bound is much tighter than the corresponding Carnot bound.

Now, we will discuss the effect of coupling the refrigerator to the cold squeezed reservoir. In the high-temperature regime,  
the mean energies at points $A$, $B$, $C$ and $D$ are given by: 
$\langle H\rangle_A= \omega_1 \coth(\beta_1\omega_1/2)\cosh(2r)/2$, 
$\langle H\rangle_B=\omega_2\lambda\coth(\beta_1\omega_1/2)\cosh(2r)/2$,
$\langle H\rangle_C=\omega_2\coth(\beta_2\omega_2/2)$,
$\langle H\rangle_D=\omega_1\lambda\coth(\beta_2\omega_2/2)/2$. The positive cooling   
condition, $\langle Q_4\rangle>0$, yields the following expressions:
\begin{widetext}
\begin{equation}
\frac{1}{2} \text{sech}(2 r)<\tau <\text{sech}(2 r)   \quad \text{and} \quad
  \zeta < \frac{3}{1-\tau  \cosh (2 r)}-2-2 \sqrt{2} \sqrt{\frac{\tau  \cosh (2 r)}{(\tau  \cosh (2 r)-1)^2}} \equiv
\zeta_{up}. \label{maxcopsq}
\end{equation}
Eq. (\ref{maxcopsq}) along with the equation (\ref{maxcop}) is our third main result. As expected, 
$\zeta_{\rm up}$ reduces to $\zeta^{\rm th}_{up}$ for the vanishing squeezing parameter, $r=0$. To discuss the physical 
significance of condition given in Eq. (\ref{maxcopsq}), we invert it in  terms of lower and upper limits on squeezing
parameter $r$:
\begin{equation}
0<\tau <\frac{1}{2},\quad \frac{1}{2} \cosh ^{-1}\left(\frac{1}{2 \tau }\right)<r<\frac{1}{2} \cosh ^{-1}\left(\frac{1}{\tau }\right) \quad\text{or}\quad \frac{1}{2}<\tau <1,\quad 0<r<\frac{1}{2} \cosh ^{-1}\left(\frac{1}{\tau }\right). \label{pcc}
\end{equation}
\end{widetext}
It is clear from the above equation that we can extract heat from squeezed cold reservoir even for $\tau<1/2$, which is
otherwise impossible with the refrigeration operation with purely thermal reservoirs. Again this can be explained on the 
basis of effective temperature of the cold reservoir [see Eq. (\ref{effectiveTemp2})]. For $r=\frac{1}{2} \cosh ^{-1}\left(\frac{1}{2 \tau }\right)$ and $r=\frac{1}{2} \cosh ^{-1}\left(\frac{1}{\tau }\right)$, the effective temperatures of the cold reservoir become $T_2/2$ and $T_2$, respectively. As per the original positive work condition ($1/2<\tau$) without cold squeezed reservoir, $T_1>T_2/2$, hence in case of cold  squeezed reservoir this condition is satisfied for the given range 
of squeezing parameter $r$ in Eq. (\ref{pcc}). Eventually, the refrigeration stops when effective temperature of cold squeezed reservoir approaches $T_2$, which is temperature of the thermal hot reservoir. Finally, for $\tau=1/2$ or $T_2=2T_1$, the allowed range of $r$ is: $0<r<\frac{1}{2} \cosh ^{-1}(2)$, which implies that effective temperature of cold reservoir should be smaller than $2T_1$ which is natural.
\section{Conclusions}

We have investigated the performance of a HQOC,  operating in the sudden switch limit, coupled to a squeezed 
thermal reservoir. First, we  showed that even in the presence of very large squeezing ($r\rightarrow\infty$), 
the maximum efficiency of the engine is 1/2 only.  This is due to the frictional effects caused 
by the non-adiabatic transitions when we operate in the sudden switch regime. Our study is in contrast with the 
previous studies which  claim that the efficiency can reach unity for large squeezing.
Then we obtained closed form expression for the upper bound on the efficiency of the engine 
operating in the high-temperature regime. The result is interesting in the sense that the obtained bound is
independent of the parameters of the model under consideration and depends on the ratio of the reservoir temperatures 
and squeezing parameter $r$ only. Additionally, we also derive the analytic expression for the efficiency at maximum work
and showed that it satisfies the derived upper bound.
As a special case of our more general setup, when squeezing parameter $r\rightarrow 0$, 
our results correspond to the case in which engine is running between two purely thermal reservoirs. 
Further, we have also obtained upper bounds for the Otto refrigerator working between two purely thermal 
reservoirs as well as for the case when cold reservoir is taken to be squeezed thermal reservoir. Finally, we showed 
that squeezing can help in cooling process otherwise impossible in standard setup with  thermal reservoirs.
\bibliography{bibloOtto}
\bibliographystyle{apsrev4-1}

\end{document}